\documentclass[aps,showpacs,preprintnumbers,amsmath, amssymb]{revtex4}

\usepackage{float}
\usepackage{graphics,epsfig}
\usepackage{graphicx}
\usepackage{dcolumn}
\usepackage{bm}

\begin{document}
\baselineskip=0.8 cm

\title{{\bf On instabilities of scalar hairy regular compact reflecting stars}}
\author{Yan Peng$^{1}$\footnote{yanpengphy@163.com}}
\affiliation{\\$^{1}$ School of Mathematical Sciences, Qufu Normal University, Qufu, Shandong 273165, China}

\vspace*{0.2cm}
\begin{abstract}
\baselineskip=0.6 cm
\begin{center}
{\bf Abstract}
\end{center}

We study the system constructed by charged scalar fields linearly coupled
to asymptotically flat horizonless compact reflecting stars.
We obtain bounds on the charge of the scalar field,
below which the scalar hairy star is expected
to suffer from nonlinear instabilities.
It means that scalar hairy regular configurations
are unstable for scalar fields of small charge.
For the highly-charged star,
there are also bounds on radii of regular compact reflecting stars.
When the star radius is below the bound,
the hairy star is always unstable.

\end{abstract}

\pacs{11.25.Tq, 04.70.Bw, 74.20.-z}\maketitle
\newpage
\vspace*{0.2cm}

\section{Introduction}

A well known characteristic of classical black holes is the
no scalar hair theorem \cite{Bekenstein,Chase,Ruffini-1},
which states that static scalar fields cannot
condense outside black holes in the asymptotically flat background, see references
\cite{Hod-1}-\cite{Brihaye} and reviews \cite{Bekenstein-1,CAR}.
And it was usually believed that this no hair property is
due to the existence of black hole horizons,
which can absorb matter fields and radiations.
So it is interesting to further examine whether this no hair behavior
is restricted to the spacetime with a horizon.

Hod firstly proved that the neutral static massive scalar field cannot condense
outside the horizonless compact reflecting stars in the asymptotically flat gravity \cite{Hod-6}.
And the asymptotically flat regular reflecting star also cannot support the
neutral massless scalar field nonminimally coupled to the gravity \cite{Hod-7}.
Including a positive cosmological constant,
it was shown that there is also no hair theorem for the neutral scalar field
in the asymptotically dS regular reflecting star spacetime \cite{Bhattacharjee}.
It means that the neutral scalar hair usually cannot form outside the reflecting star
but the system composed of charged scalar fields and charged horizonless backgrounds is still to be studied.
In fact, the regular configurations with charged scalar hair supported by a charged reflecting shell
were constructed when the shell radius is below an upper bound
and above the upper bound, the scalar field cannot exist outside the shell \cite{Hod-8,Hod-9,YP-1}.
And in the reflecting star background,
an upper bound for the radius of the charged star with charged hair was also obtained
and the no scalar hair theorem still holds in the case that the star radius
is above the upper bound \cite{Hod-10,Hod-11,Hod-12,Hod-13,YP-2,YP-3,YP-4,YP-5}.

On the other side, the general relativity predicts that
closed light rings may exist outside compact objects, such as
black holes and horizonless compact stars \cite{P1,P2,P3,P4,P5}.
It was stated that horizonless compact stars with stable closed light rings
are expected to develop nonlinear instabilities
due to the fact that massless perturbation fields tend
to pile up on stable null geodesics \cite{IS1,IS2,IS3}.
And it was shown that the innermost null circular geodesic of a horizonless
compact object is stable \cite{IN1,IN2}.
So regular compact objects with outermost light rings
above the object surface are dynamically unstable \cite{ISP}.
We should emphasize that the dynamical instability
is expected to develop for compact objects with stable null circular geodesics
which are located inside the object.
At present, various types of regular scalar hairy configurations supported by
compact reflecting stars have been constructed,
but to the best of knowledge, (in)stabilities of these scalar
hairy stars hasn't been studied. In this work, we plan to disclose
the (in)stability of the hairy reflecting star
through behaviors of light rings.

The rest of this work is planed as follows.
In section II, we introduce the system constructed by a charged
scalar field linearly coupled to a charged reflecting star.
And in section III, we obtain bounds on the star radius and scalar field charge,
which can be used to describe the instability of the hairy star.
We will summarize main results at the last section.

\section{The scalar field and reflecting star gravity system}

We are interested in the system constructed by a charged scalar field
coupled to the charged compact reflecting star
in the asymptotically flat background.
And the matter field Lagrange density reads
\begin{eqnarray}\label{lagrange-1}
\mathcal{L}=-\frac{1}{4}F^{MN}F_{MN}-|\nabla_{\mu} \psi-q A_{\mu}\psi|^{2}-m^{2}\psi^{2}.
\end{eqnarray}
Here $\psi(r)$ and $A_{\mu}$ are the scalar field and Maxwell field respectively.
We also label q as the scalar field charge and m as the scalar field mass.

The general spherically symmetric compact star solution can be expressed as
\cite{P1}
\begin{eqnarray}\label{AdSBH}
ds^{2}&=&-e^{-2\chi}fdt^{2}+\frac{dr^{2}}{f}+r^{2}(d\theta^{2}+sin^{2}\theta d\varphi^{2}).
\end{eqnarray}
where $\chi(r)$ and $f(r)$ are metric functions satisfying $\chi(r\rightarrow\infty)\rightarrow 0$
and $f(r\rightarrow\infty)\rightarrow 1$.
We define $r_{s}$ as the star radius and there is $f(r)>0$ for $r\geqslant r_{s}$
since we study the regular compact star.

According to the Einstein equations $G^{\mu}_{\nu}=8\pi T^{\mu}_{\nu}$,
the metric equations are \cite{ISP,dyp,metric1,metric2,metric3}
\begin{eqnarray}\label{BHg}
f'=-8\pi r [\rho+\frac{Q^{2}(r)}{8\pi r^4}]+(1-f)/r,
\end{eqnarray}
\begin{eqnarray}\label{BHg}
\chi'=-4\pi r (\rho+p)/f
\end{eqnarray}
with $T^{t}_{t}=-\rho$, $T^{r}_{r}=p$ and $Q(r)$
is the electric charge contained within a sphere of
area of radius r.

In the present paper, we are interested in neglecting
the scalar hair's backreaction on the metric
and there is $\chi(r)=0$ and $f(r)=1-\frac{2M}{r}+\frac{Q^2}{r^2}$
with $M$ as the ADM mass and $Q$ corresponding to the star charge.
We also assume that the Maxwell field has only the nonzero $t$
component in the form $A_{t}=-\frac{Q}{r}dt$.
And the equation of radial dependence scalar field $\psi=\psi(r)$ is
\begin{eqnarray}\label{BHg}
\psi''+(\frac{2}{r}+\frac{f'}{f})\psi'+(\frac{q^2Q^2}{r^2f^2}-\frac{m^2}{f})\psi=0
\end{eqnarray}
with $f=1-\frac{2M}{r}+\frac{Q^2}{r^2}$ \cite{YP-2,blc1,blc2,blc3,blc4,blc5}.

The compact star has a scalar reflecting surface
that the scalar field vanishes at the star radius $r_{s}$.
At the infinity, the scalar field possesses the asymptotical behavior
of $\psi\sim A\cdot \frac{1}{r}e^{-mr}+B\cdot \frac{1}{r}e^{mr}$
with A and B as integral constants.
We set $B=0$ to get the physical scalar field solution
and boundary conditions of the scalar field are
\begin{eqnarray}\label{InfBH}
&&\psi(r_{s})=0,~~~~~~~~~~~~~\psi(\infty)=0.
\end{eqnarray}

According to the general relativity, closed light rings
may exist outside compact stars.
In this work, the outer light ring of the charged hairy compact star is
$r_{\gamma}=\frac{1}{2}(3 M+\sqrt{9M^2-8Q^2})$,
which refers to the outer null circular geodesic above the surface of the compact object
where the contribution of the linearized field to the
spacetime metric is neglected \cite{ISP}.
In fact, the innermost circular light ring of a horizonless compact object is stable.
And it was stated that horizonless compact stars are
unstable if the stars possess stable closed light rings \cite{IS1,IS2,IS3}.
We realize that the former studies have shown that generic horizonless compact
objects are characterized by an even number of light rings. Thus, the presence
of a light ring outside the objects implies the presence of an inner light ring as well \cite{IN1}.
So we can disclose the stability of hairy stars
by examining whether there is the outer light ring
outside the star surface.

\section{Bounds for the charge of scalar fields supported by reflecting stars}

\subsection{Bounds of the scalar field charge in the case of $\frac{Q}{M}\leqslant1$}

Introducing a new radial function $\tilde{\psi}=\sqrt{r}\psi$,
the scalar field equation (5) is transformed into
\begin{eqnarray}\label{BHg}
r^2\tilde{\psi}''+(r+\frac{r^2f'}{f})\tilde{\psi}'+(-\frac{1}{4}-\frac{rf'}{2f}+\frac{q^2Q^2}{f^2}-\frac{m^2r^2}{f})\tilde{\psi}=0,
\end{eqnarray}
where $f=1-\frac{2M}{r}+\frac{Q^2}{r^2}$.

From the relation (6), we get the following boundary conditions
\begin{eqnarray}\label{InfBH}
&&\tilde{\psi}(r_{s})=0,~~~~~~~~~\tilde{\psi}(\infty)=0.
\end{eqnarray}

One extremum point $r=r_{peak}$ of the function $\tilde{\psi}$
exists between the star surface $r_{s}$ and the infinity boundary.
At this extremum point $r_{peak}$, the following relation holds
\begin{eqnarray}\label{InfBH}
\{ \tilde{\psi}'=0~~~~and~~~~\tilde{\psi} \tilde{\psi}''\leqslant0\}~~~~for~~~~r=r_{peak}.
\end{eqnarray}

With the relations (7) and (9), we obtain the inequality
\begin{eqnarray}\label{BHg}
-\frac{1}{4}-\frac{rf'}{2f}+\frac{q^2Q^2}{f^2}-\frac{m^2r^2}{f}\geqslant0~~~for~~~r=r_{peak}.
\end{eqnarray}

It can be transformed into
\begin{eqnarray}\label{BHg}
m^2r^2f(r)\leqslant q^2Q^2-\frac{rff'}{2}-\frac{1}{4}f^2~~~for~~~r=r_{peak}.
\end{eqnarray}

Firstly, we have $r_{s}\geqslant M+\sqrt{M^2-Q^2}$
since there is an horizon above the surface of
the compact object in the case of $ r_{s}< M+\sqrt{M^2-Q^2}$.
Considering $r_{s}\geqslant M+\sqrt{M^2-Q^2}$,
the following relations hold
\begin{eqnarray}\label{BHg}
r\geqslant r_{s}\geqslant M+\sqrt{M^2-Q^2}\geqslant M\geqslant Q,
\end{eqnarray}
\begin{eqnarray}\label{BHg}
f=1-\frac{2M}{r}+\frac{Q^2}{r^2}=\frac{1}{r^2}[(r-M)^2-(M^2-Q^2)]\geqslant0,
\end{eqnarray}
\begin{eqnarray}\label{BHg}
rf'=r(\frac{2M}{r^2}-\frac{2Q^2}{r^3})=\frac{2M}{r}(1-\frac{Q}{r}\frac{Q}{M})\geqslant 0,
\end{eqnarray}
\begin{eqnarray}\label{BHg}
(r^2f)'=(r^2-2Mr+Q^2)'=2(r-M)\geqslant 0.
\end{eqnarray}

From relations (11) and (15), $r^2f$ is an increasing function and we have
\begin{eqnarray}\label{BHg}
m^2r_{s}^2f(r_{s})\leqslant m^2r^2f(r)\leqslant q^2Q^2-\frac{rff'}{2}-\frac{1}{4}f^2\leqslant q^2Q^2~~~for~~~r=r_{peak}.
\end{eqnarray}

According to (16), there is $m^2r_{s}^2f(r_{s})\leqslant  q^2Q^2$ or
\begin{eqnarray}\label{BHg}
m^2r_{s}^2(1-\frac{2M}{r_{s}}+\frac{Q^2}{r_{s}^2})\leqslant  q^2Q^2.
\end{eqnarray}

The inequality can also be transformed into
\begin{eqnarray}\label{BHg}
(m r_{s})^2-(2m M)(m r_{s})+Q^2(m^2-q^2)\leqslant 0.
\end{eqnarray}

With the relation (18), we obtain bounds on radii of hairy stars as
\begin{eqnarray}\label{BHg}
m r_{s}\leqslant m M+\sqrt{m^{2}(M^2-Q^2)+q^2Q^2},
\end{eqnarray}
with dimensionless quantities according to the symmetry
\begin{eqnarray}\label{BHg}
r\rightarrow k r,~~~~ m\rightarrow m/k,~~~~ M\rightarrow k M,~~~~ Q\rightarrow k Q,~~~~ q\rightarrow q/k.
\end{eqnarray}

In order to obtain the instability condition,
we impose that the outer light ring is above the upper bound (19) in the form
\begin{eqnarray}\label{BHg}
m r_{\gamma}=\frac{1}{2}(3m M+\sqrt{9m^{2}M^2-8m^2Q^2})\geqslant m M+\sqrt{m^{2}(M^2-Q^2)+q^2Q^2}.
\end{eqnarray}
From (21), we get the upper bound for the charge of the scalar field as
\begin{eqnarray}\label{BHg}
(\frac{q}{m})^2\leqslant\frac{3M^2}{2Q^2}+\frac{M\sqrt{9M^2-8Q^2}}{2Q^2}-1
\end{eqnarray}
So for $\frac{Q}{M}\leqslant 1$ in this part, we find that
the hairy star is unstable for
small scalar field charge below the bound (22).

\subsection{Bounds of the scalar field charge in the case of $1<\frac{Q}{M}\leqslant\sqrt{\frac{9}{8}}$}

Now we extend the discussion of (in)stabilities of hairy reflecting stars to the
range of $1<\frac{Q}{M}\leqslant\sqrt{\frac{9}{8}}$.
In the case of $r_{s}<\frac{Q^2}{M}$, there is
\begin{eqnarray}\label{BHg}
r_{\gamma}=\frac{1}{2}(3 M+\sqrt{9M^2-8Q^2})\geqslant \frac{3M}{2}\geqslant \frac{3}{2}\sqrt{\frac{8}{9}}Q
=\sqrt{2}Q\geqslant \sqrt{\frac{9}{8}}Q\geqslant\frac{Q}{M}Q> r_{s}.
\end{eqnarray}
So the light ring $r_{\gamma}$ is above the star radius $r_{s}$ and the star is unstable.

In another case of $r_{s}\geqslant\frac{Q^2}{M}$, we have
\begin{eqnarray}\label{BHg}
Q^2-M^2\geqslant 0,
\end{eqnarray}
\begin{eqnarray}\label{BHg}
1-\frac{Q}{r}\frac{Q}{M}\geqslant 1,
\end{eqnarray}
\begin{eqnarray}\label{BHg}
r\geqslant r_{s}\geqslant\frac{Q^2}{M}\geqslant Q\geqslant M.
\end{eqnarray}

And the following relations exist
\begin{eqnarray}\label{BHg}
f=1-\frac{2M}{r}+\frac{Q^2}{r^2}=\frac{1}{r^2}[(r-M)^2+Q^2-M^2)]\geqslant0,
\end{eqnarray}
\begin{eqnarray}\label{BHg}
rf'=r(\frac{2M}{r^2}-\frac{2Q^2}{r^3})=\frac{2M}{r}(1-\frac{Q}{r}\frac{Q}{M})\geqslant 0,
\end{eqnarray}
\begin{eqnarray}\label{BHg}
(r^2f)'=(r^2-2Mr+Q^2)'=2(r-M)\geqslant 0.
\end{eqnarray}

Following approaches in part A, we again
obtain the upper bound for hairy star radius
\begin{eqnarray}\label{BHg}
m r_{s}\leqslant m M+\sqrt{m^{2}(M^2-Q^2)+q^2Q^2}.
\end{eqnarray}
The same upper bound was obtained in \cite{YP-2}
on conditions that $\frac{Q}{M}\leqslant 1$, which was also shown
in (19) of part A. Here we find that the same bound holds in the other
case of $1<\frac{Q}{M}\leqslant\sqrt{\frac{9}{8}}$.

And we also arrive at the bound on the scalar field charge
the same as (22) in the form
\begin{eqnarray}\label{BHg}
(\frac{q}{m})^2\leqslant\frac{3M^2}{2Q^2}+\frac{M\sqrt{9M^2-8Q^2}}{2Q^2}-1
\end{eqnarray}

It is well know that the neutral scalar field usually cannot exist around
reflecting stars and charge scalar fields may condense outside the reflecting
star. In this part with $1<\frac{Q}{M}\leqslant\sqrt{\frac{9}{8}}$,
we show that the hairy star is always unstable
on the condition $r_{s}<\frac{Q^2}{M}$ and when $r_{s}\geqslant\frac{Q^2}{M}$,
the hairy star is still unstable for small charge of scalar fields below the bound (31).

\section{Conclusions}

We studied the system of charged scalar fields linearly coupled to regular reflecting stars.
In a parameter range of $\frac{Q}{M}\leqslant\sqrt{\frac{9}{8}}$,
we obtained upper bounds for hairy star radii
as $m r_{s}\leqslant m M+\sqrt{m^{2}(M^2-Q^2)+q^2Q^2}$,
where m and q are the mass and charge of the scalar field respectively,
M serves as the ADM mass and Q corresponds to the star charge.
Scalar fields can condense only when the star radius is below the upper bound.
And we mainly further investigated (in)stabilities of
hairy reflecting stars. We divided the discussion into two cases
as follows.

(1)~In the first case of $\frac{Q}{M}\leqslant 1$,
the hairy star is expected to suffer from nonlinear instabilities
when the charge of the scalar field is small expressed with dimensionless quantities as
$(\frac{q}{m})^2\leqslant\frac{3M^2}{2Q^2}+\frac{M\sqrt{9M^2-8Q^2}}{2Q^2}-1$.

(2)~In the second case of $1<\frac{Q}{M}\leqslant\sqrt{\frac{9}{8}}$,
we found an upper bound for the hairy star radius
as $r_{s}<\frac{Q^2}{M}$, below which the hairy star is unstable
and when the star radius is above this bound,
the hairy star is still unstable in the case that the charge of
the scalar field satisfies
$(\frac{q}{m})^2\leqslant\frac{3M^2}{2Q^2}+\frac{M\sqrt{9M^2-8Q^2}}{2Q^2}-1$.

\begin{acknowledgments}

We would like to thank the anonymous referee for the constructive suggestions to improve the manuscript.
This work was supported by the Shandong Provincial Natural Science Foundation of China under Grant
No. ZR2018QA008.

\end{acknowledgments}

\end{document}